\documentclass[a4paper]{article}

\usepackage{jheppub}
\usepackage[english]{babel}
\usepackage{bm}

\makeatletter
\def\@fpheader{\relax}
\makeatother

\newcommand{\Dslash}{D\hspace{-.67em}/\hspace{.2em}}
\newcommand{\tr}{\mathrm{tr}}

\newcommand{\ol}[1]{\bar{#1}}
\newcommand{\pbp}{\ol{\psi}\psi}
\newcommand{\res}{0.371(20)}

\title{A precise determination of the $\pbp$ anomalous dimension in conformal gauge theories}

\author{Agostino Patella}
\affiliation{PH-TH, CERN, CH-1211 Geneva 23, Switzerland}
\emailAdd{agostino.patella@cern.ch}

\abstract{
A strategy for computing the $\pbp$ anomalous dimension at the fixed point in infrared-conformal gauge theories from lattice simulations is discussed. The method is based on the scaling of the spectral density of the Dirac operator or rather its integral, the mode number. It is relatively cheap, mainly for two reasons: (a) the mode number can be determined with quite high accuracy, and (b) the $\pbp$ anomalous dimension is extracted from a fit of several observables on the \textit{same} set of configurations (no scaling in the Lagrangian parameters is needed). As an example the $\pbp$ anomalous dimension has been computed in the $\textrm{SU}(2)$ theory with $2$ Dirac fermions in the adjoint representation of the gauge group, and has been found to be $\gamma_*=\res$. In this particular case, the proposed strategy has proved to be very robust and effective.
}

\preprint{CERN-PH-TH/2012-102}


\arxivnumber{}

\begin{document}

\maketitle

\section{Introduction}

Non-Abelian gauge theories with minimally coupled Dirac fermions generally produce spontaneous chiral symmetry breaking if the number of fermions (or \textit{flavors}) is not too large. When the number of flavors is increased,\footnote{If the number of flavours is large enough, asymptotic freedom is destroyed. Only asymptotically free theories will be considered here.} chiral symmetry is restored and the beta function develops a non-Gaussian infrared fixed point \cite{Banks:1981nn}, indicating that an approximate dilatation symmetry is realized at large distances. Gauge theories in this phase are usually referred to as \textit{infrared-conformal gauge theories} (IR-CGTs). Deformed (with relevant operators) IR-CGTs have been proposed in the past years as interesting models for strongly-coupled physics beyond the Standard Model, for instance in walking or conformal technicolor scenarios (see \cite{Yamawaki:1985zg, Holdom:1984sk, Holdom:1981rm, Appelquist:1986an, Piai:2010ma, Andersen:2011yj, Hong:2004td, Dietrich:2006cm, Luty:2004ye, Azatov:2011ht}, and references therein). The anomalous dimension $\gamma_*$ of the fermion mass operator $\pbp$ at the IR-fixed point plays a very important role in all those models. Being constrained to be in between $0$ (as in the perturbative Bank-Zaks fixed point \cite{Banks:1981nn}) and $2$ (from the unitarity bound \cite{Mack:1975je, Grinstein:2008qk}), it is required to be about $1$ for interesting models. In this regime the IR-CGT is strongly coupled and a determination of $\gamma_*$ from first principles can be obtained only by means of lattice simulations.

The theory considered in this paper is the $\textrm{SU}(2)$ gauge theory with $2$ Dirac fermions in the adjoint representation of the gauge group ($\textrm{SU}(2)+2\textrm{adj}$). It is IR-conformal as confirmed by a relatively large literature and several complementary analysis strategies \cite{Catterall:2007yx, Catterall:2008qk, DelDebbio:2008zf, DelDebbio:2008tv, DelDebbio:2009fd, DelDebbio:2010hx, DelDebbio:2010hu, Bursa:2011ru, Hietanen:2008mr, DelDebbio:2011kp, Appelquist:2011dp, Hietanen:2009az, Hietanen:2009zz, Karavirta:2011mv, Bursa:2009tj, Bursa:2009we, DeGrand:2011qd, DeGrand:2011vp, Catterall:2011zf, Catterall:2011ce, Giedt:2012rj}. Although it is not clear whether this theory will be useful for building realistic technicolor models, it represents the ideal playground for testing new analysis methods. The $\pbp$ anomalous dimension has been already estimated using different techniques. The first very rough estimate $0.05 \le \gamma_* \le 0.20 $ was published in \cite{DelDebbio:2010hu} using finite-size scaling of mesonic observables (similar to the one given in \cite{Lucini:2009an}). A somewhat larger value $\gamma_* = 0.22(6)$ was found in \cite{DelDebbio:2010hx}, fitting a power of the fermion mass to the string tension. These first estimates must be taken with a grain of salt, since the systematic errors were still not well understood at that time. Using the Schr\"odinger Functional renormalization scheme, the authors of \cite{DeGrand:2011qd} quote $0.31(6)$ and the authors of \cite{Bursa:2009we} quote $0.05 \le \gamma_* \le 0.56$. As discussed in \cite{Bursa:2009we} large systematic errors are expected with this technique, because of the difficulty in localizing the IR-fixed point. Monte Carlo Renormalization Methods \cite{Catterall:2011zf} have similar problems leading to a very wide allowed range $-0.6 \le \gamma_* \le 0.6$. A more sophisticated analysis of the finite-size scaling of mesonic masses and decay constants \cite{Giedt:2012rj} yields $\gamma_*=0.51(16)$. The central value is larger than any other determination, but the error is also quite large (about $30\%$).

The method proposed in this paper to measure $\gamma_*$ is based on the observation that the spectral density $\rho$ of the Dirac operator is a power of the eigenvalue $\omega$ . Its exponent is related to the anomalous dimension through the relation \cite{Patella:2011jr,DeGrand:2009et,DelDebbio:2010ze}
\begin{gather}
\rho(\omega) = \hat{\rho}_0 \mu^{\frac{4\gamma_*}{1+\gamma_*}} \omega^{\frac{3-\gamma_*}{1+\gamma_*}} + \dots \ ,
\label{eq:rho_conformal}
\end{gather}
at the leading order for small eigenvalues ($\mu$ is the renormalization scale and $\hat{\rho}_0$ is a dimensionless constant). When scale invariance is broken by a small mass for fermions (mass-deformed IR-conformal gauge theory, IR-mCGT), eq.~\eqref{eq:rho_conformal} is valid only in an intermediate range of eigenvalues, that is not known \textit{a priori} and must be determined empirically.

In principle this method is limited by the need to have a reliable infinite-volume extrapolation, and to be close enough to the chiral limit. However we will see that the fermion mass does not need to be too small in order to see a power law in the spectral density for a wide range of eigenvalues. This surprising empirical observation, combined with the fact that the spectral density (or rather its integral) can be computed with high accuracy, will lead to a quite precise determination of the anomalous dimension, that is anticipated to be $\gamma_* = \res$. Although this result is obtained for a particular theory, the method can be exported with no modifications to any other gauge group or matter content. A precursor of the presented strategy can be found in \cite{DelDebbio:2010ze}. While this work was being finalized, an analysis of the Dirac spectral density in the $\textrm{SU}(3)$ theory with $12$ fermions in the fundamental representation appeared in~\cite{Cheng:2011ic}, although very few eigenvalues were considered there. In this work the $50$ lowest eigenvalues have been discarded on a $64 \times 24^3$ lattice, and about $2000$ eigenvalues have been considered in order to extract the $\pbp$ anomalous dimension.

\section{Spectral density}
\label{sec:spectral}

The validity of eq.~\eqref{eq:rho_conformal} is discussed in this introductory section. The eigenvalue density of the massless Dirac operator $\Dslash$ (or \textit{Dirac spectral density}) is defined as
\begin{gather}
\rho(\omega) = \lim_{V \to \infty} \frac{1}{V} \sum_k \langle \delta( \omega - \omega_k[A] ) \rangle \ ,
\end{gather}
where $i \omega_k[A] $ is the $k$-th eigenvalue of $\Dslash$ at fixed gauge configuration, in a finite box of volume $V$. At the root of eq.~\eqref{eq:rho_conformal} is the fact proved in~\cite{Giusti:2008vb}, that both the eigenvalue $\omega$ and the Dirac spectral density $\rho(\omega)$ renormalize multiplicatively. If $\gamma(g)$ is the $\pbp$ anomalous dimension at a given coupling $g$, the Dirac spectral density $\rho(\omega)$ has anomalous dimension $\gamma(g)$, while the eigenvalue $\omega$ has anomalous dimension $-\gamma(g)$.

An IR-CGT is characterised by the existence of an IR-fixed point for the running coupling $g=g_*$. The leading behaviour in eq.~\eqref{eq:rho_conformal} is obtained from the renormalization-group equation exactly at the fixed point \cite{Patella:2011jr,DeGrand:2009et,DelDebbio:2010ze}:
\begin{gather}
\left[ (1+\gamma_*) \omega \frac{\partial}{\partial \omega} - 3 + \gamma_* \right] \rho(\omega) = 0 \ ,
\label{eq:rg}
\end{gather}
which depends only on the $\pbp$ anomalous dimension $\gamma_*=\gamma(g_*)$ at the IR-fixed point. However IR-CGTs are asymptotically free at high energies and reach the IR-fixed point only at asymptotically low energies. Hence eq.~\eqref{eq:rg} and its solution~\eqref{eq:rho_conformal} are valid only for low eigenvalues.

A mass term for fermions destroys scale invariance. If the mass is small enough, the renormalization flow starts from the vicinity of the Gaussian fixed point at very high energy, moves towards the IR-fixed point, stays in its vicinity for some time, and is eventually driven away from it by the mass term. A smaller mass corresponds to a longer time spent in the vicinity of the IR-fixed point. As long as the renormalization flow stays close to the IR-fixed point, approximate scale invariance is generated.

Therefore in IR-mCGTs one can identify three regions for the eigenvalue density:
\begin{enumerate}
\item \textbf{An intermediate region of eigenvalues $\bm{\ol{\omega}_{IR} < \omega < \ol{\omega}_{UV}}$}. This region is dominated by the vicinity of the IR-fixed point. \textit{The power law~\eqref{eq:rho_conformal} is valid only in this intermediate range of eigenvalues.}
\item \textbf{A region of high eigenvalues $\bm{\omega > \ol{\omega}_{UV}}$}. Around $\omega \simeq \ol{\omega}_{UV}$ the spectral density begins its transition to the region dominated by asymptotic freedom, which will eventually lead to its asymptotic form $\rho(\omega) \propto \omega^3$. At these large eigenvalues, the spectral density is essentially insensitive to the fermion mass, hence $\ol{\omega}_{UV}$ has a well-defined nonvanishing chiral limit.
\item \textbf{A region of low eigenvalues $\bm{0 < \omega< \ol{\omega}_{IR}}$}. The spectral density feels the fermion mass  and is determined by the details of the dynamics of the IR-mCGT. No analytical model is available here \textit{a priori}. This region disappears in the chiral limit ($\ol{\omega}_{IR}$ goes to zero). Since at these low energies the physics decouples from the UV scale determining the transition to the asymptotically free regime, and since eigenvalues have the same scaling dimension of the mass, the IR scale $\ol{\omega}_{IR}$ must be proportional to $m$:
\begin{gather}
\ol{\omega}_{IR} \propto m
\label{eq:omegaIR}
\end{gather}
at the leading order in the chiral limit.
\end{enumerate}

\section{Method}
\label{sec:method}

The proposed method to extract the $\pbp$ anomalous dimension from lattice simulations is described in this section. Although the Euclidean Dirac operator is diagonalizable in the continuum theory, it is not generally so on the lattice, and one can more conveniently study the positive-definite operator:
\begin{gather}
M = (\Dslash + m)^\dag (\Dslash + m) = m^2 - \Dslash^2 \ .
\end{gather}
Following~\cite{Giusti:2008vb}, the mode number per unit volume $\bar{\nu}(\Omega)$ is defined as the number of eigenvalues of $M$ lower than $\Omega^2$ divided by the volume:
\begin{gather}
\bar{\nu}(\Omega) = 2 \int_0^{\sqrt{\Omega^2-m^2}} \rho(\omega) \ d\omega \ .
\end{gather}
In an intermediate region $\ol{\omega}_{IR}^2+m^2 < \Omega^2 < \ol{\omega}_{UV}^2 + m^2$, the mode number per unit volume is:
\begin{gather}
\bar{\nu}(\Omega)
= 2 \int_0^{\ol{\omega}_{IR}} \rho(\omega) \ d\omega + 2 \int_{\ol{\omega}_{IR}}^{\sqrt{\Omega^2-m^2}} \rho(\omega) \ d\omega \ .
\end{gather}
The first term  will be named $\bar{\nu}_0(m)$. It is an unknown additive constant, independent of $\Omega$, that depends only on the mass $m$ (remember that $\ol{\omega}_{IR} \propto m$). As discussed in sec.~\ref{sec:spectral}, in the second integral the spectral density can be approximatively replaced by the power law in eq.~\eqref{eq:rho_conformal}. Putting it all together:
\begin{gather}
\bar{\nu}(\Omega)
\simeq
\bar{\nu}_0(m) + \frac{1}{2} (1+\gamma_*) \hat{\rho}_0 \mu^{\frac{4\gamma_*}{1+\gamma_*}} (\Omega^2-m^2)^{\frac{2}{1+\gamma_*}} \ .
\label{eq:nu_powerlaw}
\end{gather}
Notice that the prefactor and the exponent of the power law do not depend on the fermion mass, while the additive constant $\bar{\nu}_0$ does (and, in particular, vanishes in the chiral limit). Since the mode number per unit volume is renormalization-group invariant~\cite{Giusti:2008vb} and has the dimension of a mass to the $4$th power, the additive constant must scale like
\begin{gather}
\bar{\nu}_0(m) \propto M_{PS}^4 \ ,
\label{eq:nu0_scaling}
\end{gather}
where $M_{PS}$ is the mass of the isotriplet pseudoscalar meson (or any other particle mass).

All formulae written so far are expressed in terms of renormalized quantities. In order to compare eq.~\eqref{eq:nu_powerlaw} to lattice data, one has to trade the renormalized quantities with the bare ones. This can be done by replacing (see~\cite{Giusti:2008vb} for details)
\begin{gather}
\Omega \to \frac{\Omega}{Z_P} \ ,
\quad
m \to \frac{m}{Z_P} = \frac{Z_A}{Z_P} m_{PCAC} \ ,
\label{eq:ren_to_bare}
\end{gather}
where $Z_A$ and $Z_P$ are the renormalization constants of the isovector axial current $A_\mu$ and the isovector pseudoscalar density $P^a$ respectively, and $m_{PCAC}$ is the quark mass as defined from the \textit{bare} PCAC (partially-conserved axial current) relation
\begin{gather}
\partial_\mu A^a_\mu = m_{PCAC} P^a \ .
\end{gather}
The mode number does not need to be redefined, since it is renormalization-group invariant. Putting everything together, the mode number per unit volume in lattice units as a function of the \textit{bare} eigenvalue $\Omega$ becomes
\begin{gather}
a^{-4} \bar{\nu}(\Omega)
\simeq
a^{-4} \bar{\nu}_0 + A [(a\Omega)^2-(am)^2]^{\frac{2}{1+\gamma_*}} \ ,
\label{eq:nu_powerlaw_lattice}
\end{gather}
for a suitable definition of the dimensionless constant $A$. The parameter $m$ appearing in the previous formula is related to the \textit{bare} PCAC mass through the relation
\begin{gather}
m = Z_A m_{PCAC} \ .
\label{eq:m_to_pcac}
\end{gather}

\textit{
The proposed strategy consists in computing the mode number per unit volume by means of lattice simulations, and in extracting the $\pbp$ anomalous dimension by fitting eq.~\eqref{eq:nu_powerlaw_lattice} to the lattice data. Since the validity range $\ol{\Omega}_{IR} < \Omega < \ol{\Omega}_{UV}$ of eq.~\eqref{eq:nu_powerlaw_lattice} is not known \textit{a priori}, it has to be determined by studying quality and stability of the fit procedure.
}

The mode number per unit volume can be computed essentially in two ways.
\begin{enumerate}
\item The eigenvalues of $M$ can be explicitly computed starting from the lowest one using the Chebyshev accelerated subspace iteration method described in detail in \cite{DelDebbio:2005qa}. Since the number of eigenvalues below some fixed value $\Omega^2$ grows linearly with the volume, this method is unpractical for large volumes.
\item Alternatively (following ref.~\cite{Giusti:2008vb}) one can define the projector $\mathbb{P}(\Omega)$ over the eigenspaces of $M$ corresponding to lower eigenvalues than $\Omega^2$, in terms of which the mode number per unit volume is
\begin{gather}
\bar{\nu}(\Omega) = \lim_{V \to \infty} \frac{1}{V} \langle \tr \ \mathbb{P}(\Omega) \rangle \ .
\label{eq:nu_p}
\end{gather}
The projector can be approximated by a suitable rational function:
\begin{gather}
\mathbb{P}(\Omega) \simeq h(\mathbb{X})^4 , \qquad \mathbb{X} = 1- \frac{2 \Omega_*^2}{M + \Omega_*^2} ,
\label{eq:p_approx}
\end{gather}
where $h(x)$ is a polynomial, and $\Omega_*$ is a parameter of order $\Omega$ defined in eq.~\eqref{eq:omegastar}. The trace in eq.~\eqref{eq:nu_p} can be estimated stochastically. The error due to the approximation in eq.~\eqref{eq:p_approx} can be estimated \textit{a posteriori}, once the spectral density is reconstructed by means of the fit procedure (see appendix \ref{app:error}). This technique is more effective for larger volumes, and is described in detail in \cite{Giusti:2008vb}.
\end{enumerate}

\section{Analysis and results}

\begin{table}[t]
\centering
\begin{tabular}{|c||c|c||c|c||c|}
\hline
Set & lattice & $am_0$ & $a m_{PCAC}$ & $aM_{PS}$ & References \\
\hline
\hline
S1 & $64\times 24^3$ & $-1.15$ & $0.11844(63)$ & $0.6414(41)$ & \cite{DelDebbio:2011kp} \\
\hline
S2 & $64\times 32^3$ & $-1.15$ & $0.11790(36)$ & $0.6386(15)$ & \cite{DelDebbio:2011kp} \\
\hline
S3 & $64\times 24^3$ & $-1.18$ & $0.05507(89)$ & $0.3374(63)$ & \cite{Bursa:2011ru} \\
\hline
\end{tabular}
\caption{Summary of the three sets, used for computing the mode number. The PCAC and isotriplet pseudoscalar meson masses have been published in the cited papers, and are given here just for the reader's convenience.}
\label{tab:sets}
\end{table}

In order to illustrate the proposed technique, let us consider the $\textrm{SU}(2)+2\textrm{adj}$ theory, which is in the conformal window.

Numerical simulations have been performed at fixed $\beta=2.25$ with Wilson fermions. I have used three sets (table~\ref{tab:sets}) of configurations generated with the HiRep code \cite{DelDebbio:2008zf} and already used in previous works to compute the mass spectrum (mesons, glueballs and string tension) \cite{DelDebbio:2009fd, DelDebbio:2010hx, DelDebbio:2010hu, Bursa:2011ru, DelDebbio:2011kp}. The first set (S1) corresponds to bare mass $am_0=-1.15$ on a $64\times 24^3$ lattice, and includes 25 well-decorrelated configurations. The second set (S2) corresponds to the same bare mass $am_0=-1.15$ but on the larger $64\times 32^3$ lattice, and includes 20 well-decorrelated configurations. The third set (S3) corresponds to a lighter bare mass $am_0=-1.18$ on a $64 \times 24^3$ lattice and includes 20 well-decorrelated configurations.

The $\pbp$ anomalous dimension will be extracted by fitting eq.~\eqref{eq:nu_powerlaw_lattice} to the mode number per unit volume obtained from sets S1 and S3. Set S2 will be used to check that the finite-volume effects are under control at least for the higher mass. Moreover the stability of the fitting procedure will be checked against the introduction of a subleading term in eq.~\eqref{eq:nu_powerlaw_lattice}.

Before moving to the details of the analysis, the results of the following subsections are anticipated and summarized in table~\ref{tab:determinations} for the reader's convenience. The $\pbp$ anomalous dimensions in table~\ref{tab:determinations} are all compatible; however, the first one $\gamma_* = 0.371(20)$ obtained from set S1 is quoted as the final results. This choice is dictated by the fact that the finite-volume effects have been explicitly checked for set S1. Even considering the weighted average between the two determinations obtained in the same range $0.091 \le a\Omega \le 0.18$, one would get the very similar result $\gamma_* = 0.365(19)$.

\begin{table}[!h]
\centering
\begin{tabular}{|c||c|c|c||c|}
\hline
Subsection & Data set & Fitting function & Fit range & $\gamma_*$  \\
\hline
\hline
\ref{sec:S1a} & S1 & \eqref{eq:nu_powerlaw_lattice} & $0.091 \le a\Omega \le 0.18$ & $0.371(20)$ \\
\hline
\ref{sec:S1b} & S1 & \eqref{eq:nu_powerlaw2_lattice} & $0.091 \le a\Omega \le 0.29$ & $0.355(23)$ \\
\hline
\ref{sec:S3} & S3 & \eqref{eq:nu_powerlaw_lattice} & $0.091 \le a\Omega \le 0.16$ & $0.364(92)$ \\
\hline
\ref{sec:S3} & S3 & \eqref{eq:nu_powerlaw_lattice} & $0.091 \le a\Omega \le 0.18$ & $0.325(50)$ \\
\hline
\end{tabular}
\caption{Determinations of the $\pbp$ anomalous dimension from: \textit{(a)} set S1 ($am_0=-1.15$); \textit{(b)} set S1 ($am_0=-1.15$) including a subleading correction; and \textit{(c)},\textit{(d)} set S3 ($am_0=-1.18$) with two different fit ranges. The three determinations are compatible, in the sense that the $1\sigma$ regions overlap. The first determination is quoted as the final result.}
\label{tab:determinations}
\end{table}

\subsection[Set S1: determination of gamma]{Set S1: determination of $\gamma_*$}
\label{sec:S1a}

The lowest $200$ eigenvalues of $M$ are explicitly computed. This allows us to reconstruct the mode number per unit volume up to $a\Omega = 0.1604$. Above this eigenvalue, the projector method is used. Results are listed in table~\ref{tab:S1}.

For $a\Omega \simeq 10$ the mode number per unit volume saturates at the value $12 a^{-4}$ ($12$ is the dimension of the vector space spanned by the pseudofermions in a single site). The saturation is a pure lattice artifact, so only values of $a\Omega$ for which the mode number per unit volume is less than $0.12 a^{-4}$ are considered (one hundredth of the total number of eigenvalues).

Eq.~\eqref{eq:nu_powerlaw_lattice} is fitted to the lattice data. Although the PCAC mass is known, but no determination of $Z_A$ is available, the parameter $am$ which determines the mode-number gap can not be reconstructed from eq.~\eqref{eq:m_to_pcac} and is considered as a fitting parameter along with $a^{-4} \bar{\nu}_0$, $A$ and $\gamma_*$. From the discussion in sec.~\ref{sec:method} it is also clear that the fitting function in eq.~\eqref{eq:nu_powerlaw_lattice} can be used only in an intermediate range of eigenvalues, that must be determined by studying the stability and quality of the fit.

\smallskip

\begin{figure}[ht]
\centering
\includegraphics[width=\textwidth]{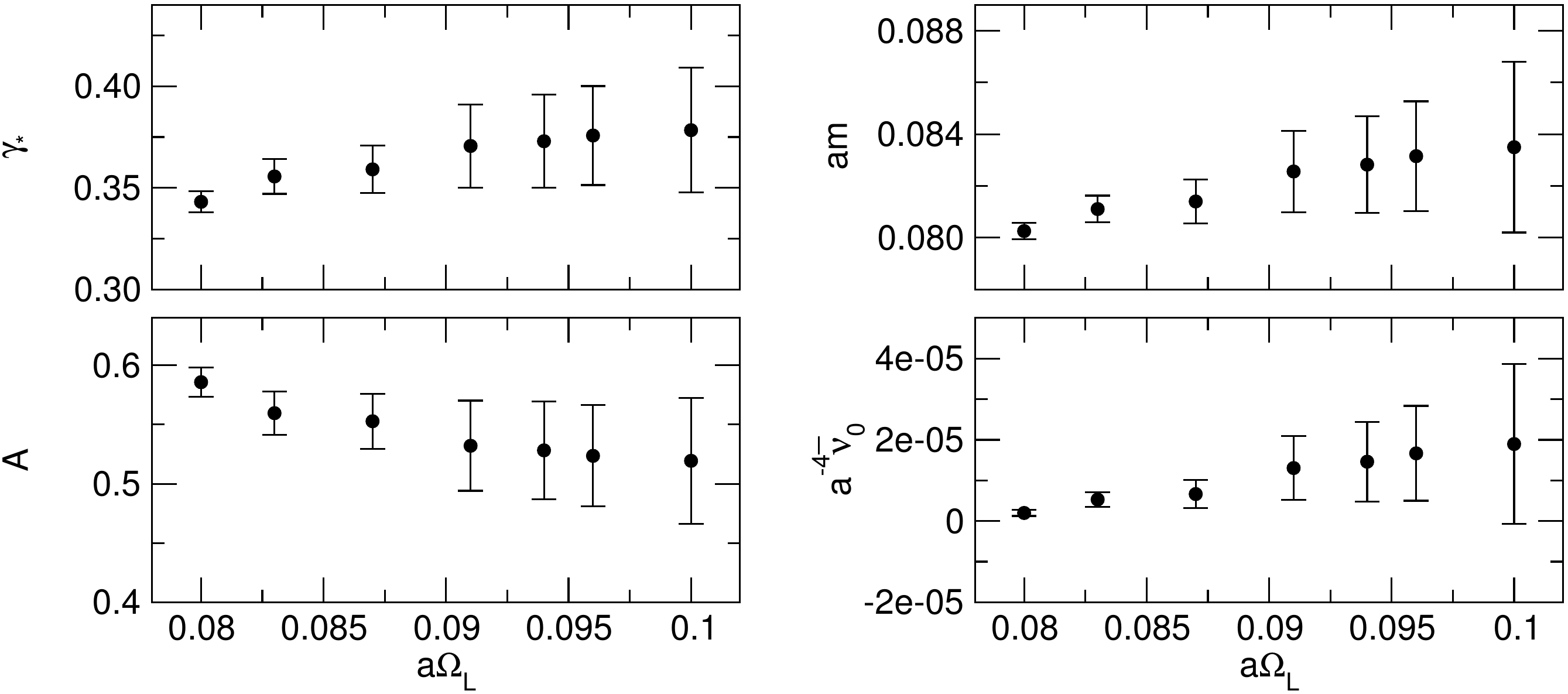}
\caption{Parameters in eq.~\eqref{eq:nu_powerlaw_lattice}, fitted to set S1 in the range $a\Omega_L \le a\Omega \le 0.18$ for various values of the lower end $a\Omega_L$. While the fit range is reduced the fit parameters shift, and reach plateaux for $a\Omega_L \simeq 0.091$.}
\label{fig:left}
\end{figure}

\textbf{Determination of the fit-range lower end.}
The fit works well with a $\chi^2/\textrm{dof}$ of order $1$ in the quite large range $0.08 \le a\Omega \le 0.4$. Choosing the fit-range higher end to be $0.18$, the lower end is systematically increased from $0.08$ to $0.1$ (fits S1:F1 to S1:F7 in table~\ref{tab:fits}). The fit parameters are shown as functions of the fit-range lower end in fig.~\ref{fig:left}. Increasing the fit-range lower end, the fit parameters keep shifting up to $a\Omega_H \simeq 0.91$ where they reach a plateau. The latter value is chosen as the fit-range lower end. Notice that at the lower end of the fit range $\bar{\nu}(0.092/a) \gtrsim 3 \bar{\nu}_0$, and the additive constant becomes soon negligible for larger eigenvalues.

\smallskip

\begin{figure}[ht]
\centering
\includegraphics[width=.8\textwidth]{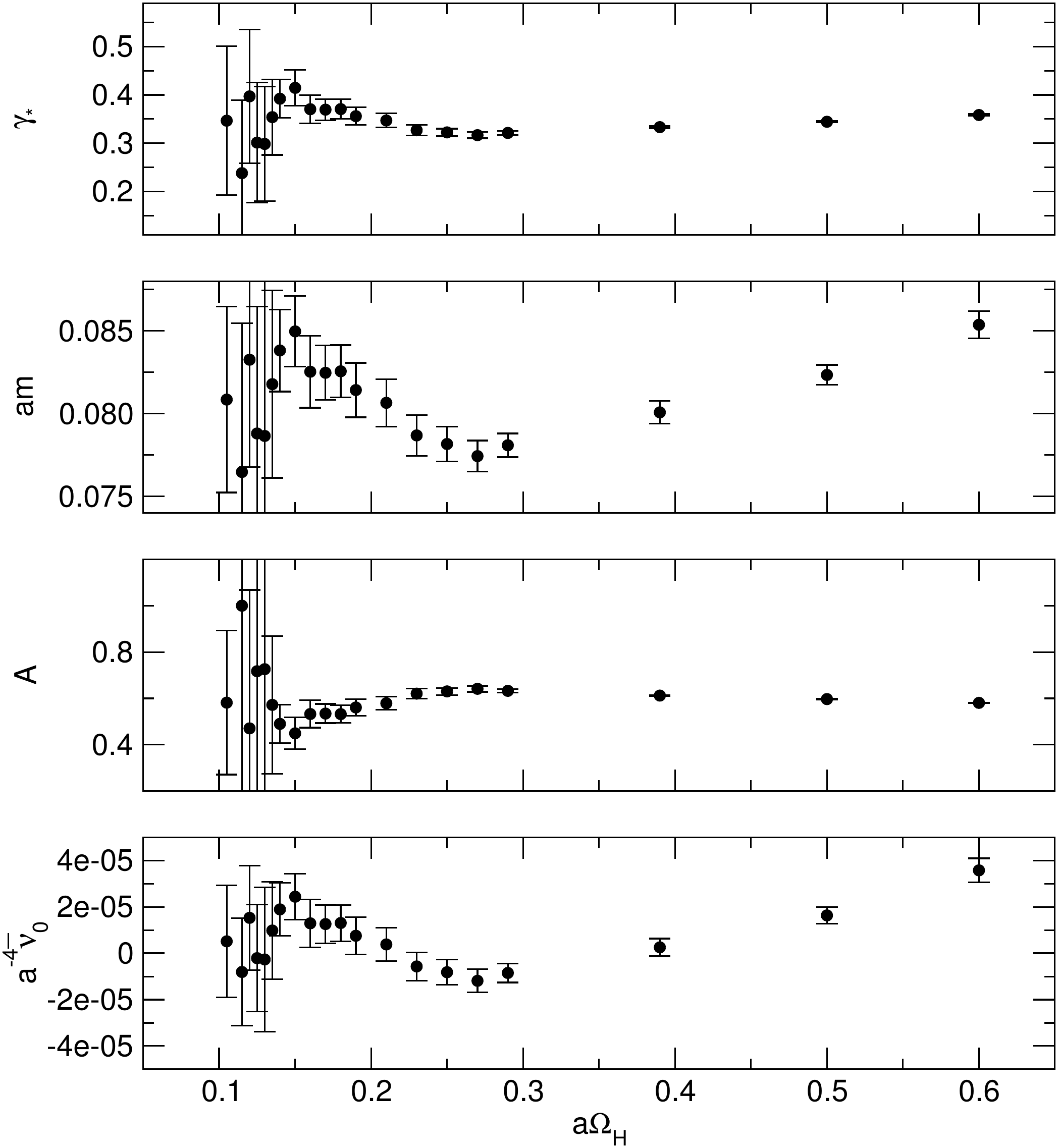}
\caption{Parameters in eq.~\eqref{eq:nu_powerlaw_lattice}, fitted to set S1 in the range $0.091 \le a\Omega \le  a\Omega_H$ for various values of the higher end $a\Omega_H$. While the fit range is reduced the fit parameters shift, and reach plateaux for $a\Omega_H \simeq 0.18$.}
\label{fig:right}
\end{figure}

\textbf{Determination of the fit-range higher end.}
The fit-range higher end is also systematically investigated. Eq.~\eqref{eq:nu_powerlaw_lattice} is fitted to the data in the range $0.091 \le a\Omega \le a\Omega_H$, where the higher end $a\Omega_H$ is gradually lowered from $0.6$ to $0.1$ (fits S1:F8 to S1:F26 in table~\ref{tab:fits}). Although the $\chi^2/\textrm{dof}$ becomes less than $1$ at about $a\Omega_H \simeq 0.4$, the fit parameters keep shifting up to $a\Omega_H \simeq 0.18$ where they reach a plateau (see plots in fig.~\ref{fig:right}). The latter value is chosen as the fit-range higher end.

\smallskip

\begin{figure}[ht]
\centering
\includegraphics[width=\textwidth]{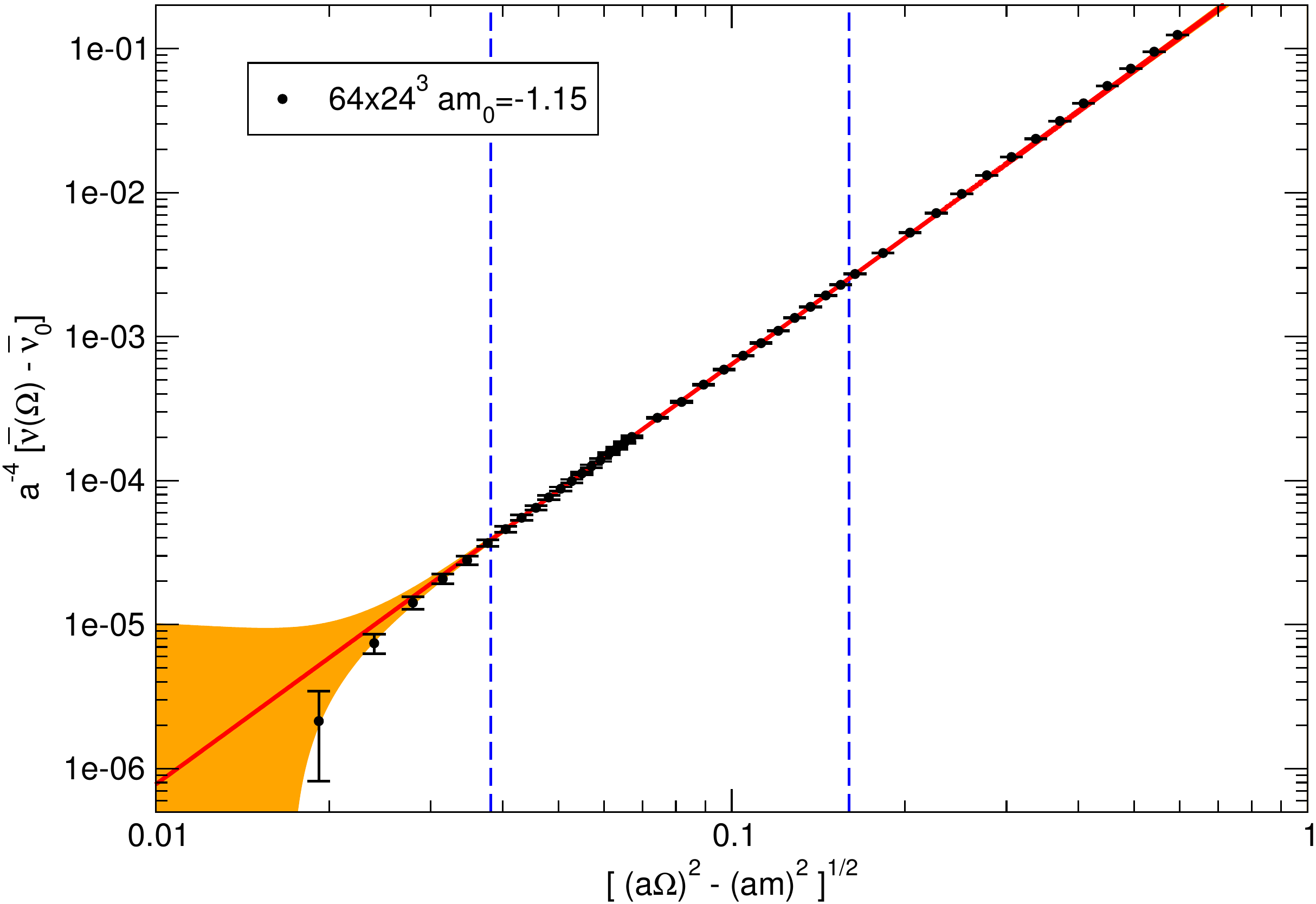}
\caption{Mode number per unit volume for the set S1 ($am_0=-1.15$ on a $64 \times 24^3$ lattice): lattice data and fit result in log-log scale. The reference fit is S1:F4 in table~\ref{tab:fits}. The parameters in the axis labels have been chosen to be $a^{-4}\bar{\nu}_0 = 1.31 \times 10^{-5}$ and $a m = 0.0826$ (best-fit results). The black points are the data computed by numerical simulations. The red line is the best fit to eq.~\eqref{eq:nu_powerlaw_lattice}, while the orange band corresponds to the $1\sigma$ region. The blue dashed lines delimit the data used for the fit.}
\label{fig:nu}
\end{figure}

Summarizing, the chosen fit range is $0.091 \le a\Omega \le 0.18$ (fit S1:F4 in table~\ref{tab:fits}), which yields a mode number per unit volume of the form (fig.~\ref{fig:nu})
\begin{gather}
a^{-4} \bar{\nu}(\Omega)
\simeq
1.31(78) \times 10^{-5} + 0.532(38) [(a\Omega)^2-0.0826(16)^2]^{\frac{2}{1+0.371(20)}} \ .
\label{eq:higher_nu}
\end{gather}
The $\pbp$ anomalous dimension is determined to be $\gamma_* = 0.371(20)$. Notice that the chosen range fit corresponds to about $2000$ eigenvalues on the $64\times 24^3$ lattice.

\subsection{Set S1: corrections to the leading power}
\label{sec:S1b}

It is interesting to check the stability of the fit result~\eqref{eq:higher_nu} against the introduction of a subleading power-law contribution in eq.~\eqref{eq:nu_powerlaw_lattice}, which becomes
\begin{gather}
a^{-4} \bar{\nu}(\Omega)
\simeq
a^{-4} \bar{\nu}_0 + A [(a\Omega)^2-(am)^2]^{\frac{2}{1+\gamma_*}} + B [(a\Omega)^2-(am)^2]^\beta \ .
\label{eq:nu_powerlaw2_lattice}
\end{gather}
Since eq.~\eqref{eq:higher_nu} describes very well the data in the range $0.091 \le a\Omega \le 0.18$, one needs to enlarge the fit range in order to discriminate the subleading contribution. Fitting the data to eq.~\eqref{eq:nu_powerlaw2_lattice} in the range $0.091 \le a\Omega \le 0.29$, one obtains
\begin{flalign}
a^{-4} \bar{\nu}(\Omega)
\simeq
-0.06(68) \times 10^{-5}
&+ 0.529(49) [(a\Omega)^2-0.0798(14)^2]^{\frac{2}{1+0.355(23)}} + \nonumber \\
& + .16(11) [(a\Omega)^2-0.0798(14)^2]^{.926(90)} \ .
\end{flalign}
Notice that by including the subleading contribution, the estimate for the $\pbp$ anomalous dimension in this range has increased from $\gamma_*=0.3212(42)$ to $\gamma_*=0.355(23)$, becoming compatible with the determination of the previous subsection (in the sense that the $1\sigma$ regions overlap).

\subsection{Set S2: finite-volume effects}
\label{sec:S2}

As analyzed in \cite{DelDebbio:2011kp}, meson masses computed on the set S1 ($am_0 = -1.15$ on $64 \times 24^3$) are identical to the ones computed on the set S2 ($am_0 = -1.15$ on $64 \times 32^3$), within the statistical errors that are of the order of $0.5\%$. It is reasonable to expect that finite-volume effects are under control also for the mode number. However this is explicitly checked by computing the mode number per unit volume using the projector method for few values of $a\Omega$. The agreement is always within $1\sigma$ as shown in table~\ref{tab:finitevolume}. Since larger finite-volume effects are expected for lower eigenvalues, we can conclude that the finite-volume effects for the set S1 are always negligible with respect to the statistical errors for $a\Omega \ge 0.086$.

\subsection{Set S3: lighter mass}
\label{sec:S3}

The set S3 ($am_0 = -1.18$ on $64 \times 24^3$) is used to check the stability of the $\pbp$ anomalous dimension while going closer to the chiral limit. For this set no detailed investigation of finite-volume effects is available. However the isotriplet pseudoscalar meson is expected to be about $10\%$ lighter than in infinite volume (see analysis in \cite{DelDebbio:2011kp}). Similarly one has to expect sizable finite-volume effects also for the spectral density at low eigenvalues, while for larger eigenvalues the finite-volume effects become smaller. I will work under the assumption that the finite-volume effects are comparable in the two sets S1 and S3 at fixed eigenvalue. Therefore the analysis is restricted to the safe range $a \Omega \ge 0.086$.

The mode number per unit volume for this set has been computed using the projector method, and results are listed in table~\ref{tab:S3}. Eq.~\eqref{eq:nu_powerlaw_lattice} has been fitted to the data in the range $0.091 \le a\Omega \le a\Omega_H$, for several values of the right end $a\Omega_H$ (fits S3:F1 to S3:F5 in table~\ref{tab:fits}). All fits have $\chi^2/\textrm{dof}$ of order $1$, and give compatible results (the $1\sigma$ regions overlap). The value of the $\pbp$ anomalous dimension obtained from the largest fit range is $\gamma_* = 0.325(50)$ (fit S3:F5 in table~\ref{tab:fits}). Both the amplitude $A$ and the anomalous dimension $\gamma_*$, that are mass-independent, are compatible with the ones obtained from the heavier mass (fit S1:F4 in table~\ref{tab:fits}), in the sense that the $1\sigma$ regions overlap.

\section{Conclusions}

The $\pbp$ anomalous dimension of the $\textrm{SU}(2)$ gauge theory with $2$ Dirac fermions in the adjoint representation ($\textrm{SU}(2)+2\textrm{adj}$) is extracted from the mode number per unit volume of the operator $M=m^2 - \Dslash^2$, which is related to the Dirac spectral density $\rho(\omega)$ by
\begin{gather}
\bar{\nu}(\Omega) = 2 \int_0^{\sqrt{\Omega^2-m^2}} \rho(\omega) \ d\omega \ .
\end{gather}
The mode number per unit volume is computed by means of lattice simulations, using the methods described in~\cite{DelDebbio:2005qa} and~\cite{Giusti:2008vb}.

In an IR-conformal gauge theory (IR-CGT), the spectral density at small eigenvalues follows a power law, and so does the mode number. The exponent is related to the $\pbp$ anomalous dimension $\gamma_*$ at the IR-fixed point. If conformality is explicitly broken by a mass for the fermions, the spectral density is expected to follow the power law only in an intermediate range of eigenvalues. In this intermediate region, the mode number per unit volume (in lattice units) has approximatively the form
\begin{gather}
a^{-4} \bar{\nu}(\Omega)
\simeq
a^{-4} \bar{\nu}_0 + A [(a\Omega)^2-(am)^2]^{\frac{2}{1+\gamma_*}} \ .
\end{gather}

This work proves that it is possible to use the previous formula to extract the $\pbp$ anomalous dimension from lattice simulations. The strength of this method relies on the following facts:
\begin{itemize}
\item The mode number can be obtained with quite high accuracy through lattice simulations, even with few configurations ($25$ configurations have been used for the higher mass), since it is an extensive quantity. Its computation does not involve fitting procedures (like for particle masses).
\item Since the finite-volume effects on the mode number are smaller for larger eigenvalues, useful information can be extracted also from not-so-large volumes.
\item The $\pbp$ anomalous dimension is extracted from different observables (i.e. the mode number at different eigenvalues) computed on the same set of configurations. All the other known methods are based on the analysis of the scaling of a fixed observable with some parameters in the action (typically the mass, the volume or both).
\end{itemize}
It is also surprising to observe that the region controlled by the power law is already quite wide at some intermediate mass ($a M_{PS} \simeq 0.5$). This last observation has an empirical nature and is in principle model-dependent.

The $\pbp$ anomalous dimension of the $\textrm{SU}(2)+2\textrm{adj}$ is found to be $\gamma_* = \res$ at fixed $\beta=2.25$. The stability of this result has been checked by lowering the fermion mass and by including a subleading power in the fitting function. An analysis of the continuum limit will be attempted in the future. It is worth recalling again that, even though the presented result is obtained for a particular theory, the method can be exported with no modifications to any other gauge group or matter content.

\acknowledgments

I thank Martin L\"uscher for very useful discussions. I am grateful to Luigi Del Debbio, Biagio Lucini, Claudio Pica and Antonio Rago for letting me use the configurations that I have analyzed in this work. A special acknowledgment goes to Antonio Rago for his crucial help in the last stage of this work.

\appendix

\section{Projector-approximation error estimate}
\label{app:error}

It is worth recalling a  few essential facts about the procedure for the approximation of the projector $\mathbb{P}(\Omega)$ (for details the reader should refer to \cite{Giusti:2008vb}). Let $P(x)$ be the min-max polynomial of degree $n$ that minimizes the deviation
\begin{gather}
\delta = \max_{\epsilon \le y \le 1} | 1 - \sqrt{y} P(y) | \ ,
\end{gather}
and define the function
\begin{gather}
h(x) = \frac{1}{2} \left[ 1 - x P(x^2) \right] \ .
\end{gather}
The polynomial $P(x)$ approximates the function $x^{-1/2}$ in the range $\epsilon \le x \le 1$, whilst $h(x)$ approximates the function $\theta(-x)$ in the range $\sqrt{\epsilon} \le |x| \le 1$. Two degrees of approximation have been used in this paper (depending on the required precision): \textit{(a)} a polynomial of degree $32$, with $\epsilon = 10^{-2}$ and $\delta \simeq 4.35 \times 10^{-4}$; \textit{(b)} a polynomial of degree $100$, with $\epsilon = 10^{-3}$ and $\delta \simeq 5.20 \times 10^{-4}$.

The error due to the approximation of the projector $\mathbb{P}(\Omega)$ with the rational function $h(\mathbb{X})^4$ has the following spectral representation:
\begin{gather}
\Delta = \int_0^\infty d\omega \ \left[ \theta(\Omega-\omega) - h(x_\omega) \right] a^{-4} \bar{\nu}'(\omega) \ ,
\label{eq:error}
\\
x_\omega = 1 - \frac{2\Omega_*}{\omega^2+\Omega_*} \ .
\end{gather}

The quantity $\Omega_*$ is defined as
\begin{gather}
\frac{\Omega}{\Omega_*} = \left( \frac{1-\sqrt{\epsilon}}{1+\sqrt{\epsilon}} \right)^{1/2} +
\int_{-\sqrt{\epsilon}}^{\sqrt{\epsilon}} dx \ \frac{1+x}{(1-x^2)^{3/2}} h(x)^4 \ .
\label{eq:omegastar}
\end{gather}

The error $\Delta$ in eq.~\eqref{eq:error} has been estimated by using the functional form of the mode number per unit volume obtained from the best fit:
\begin{gather}
a^{-4} \bar{\nu}(\Omega)
\simeq
\begin{cases}
a^{-4} \bar{\nu}_0 + 0.532 [(a\Omega)^2-0.0826^2]^{\frac{2}{1.371}} & \textrm{for } am_0=-1.15 \\
a^{-4} \bar{\nu}_0 + 0.667 [(a\Omega)^2-0.048^2]^{\frac{2}{1.325}} & \textrm{for } am_0=-1.18 \\
\end{cases}
\ .
\end{gather}
In all cases the error $\Delta$ has been checked to be smaller that the statistical error on $a^{-4} \bar{\nu}(\Omega)$.

\bibliography{dirac}

\providecommand{\href}[2]{#2}\begingroup\raggedright\begin{thebibliography}{10}

\bibitem{Banks:1981nn}
T.~Banks and A.~Zaks, {\it {On the Phase Structure of Vector-Like Gauge
  Theories with Massless Fermions}},  {\em Nucl. Phys.} {\bf B196} (1982) 189.

\bibitem{Yamawaki:1985zg}
K.~Yamawaki, M.~Bando, and K.-i. Matumoto, {\it {Scale Invariant Technicolor
  Model and a Technidilaton}},  {\em Phys. Rev. Lett.} {\bf 56} (1986) 1335.

\bibitem{Holdom:1984sk}
B.~Holdom, {\it {Techniodor}},  {\em Phys. Lett.} {\bf B150} (1985) 301.

\bibitem{Holdom:1981rm}
B.~Holdom, {\it {Raising the Sideways Scale}},  {\em Phys. Rev.} {\bf D24}
  (1981) 1441.

\bibitem{Appelquist:1986an}
T.~W. Appelquist, D.~Karabali, and L.~C.~R. Wijewardhana, {\it {Chiral
  Hierarchies and the Flavor Changing Neutral Current Problem in Technicolor}},
   {\em Phys. Rev. Lett.} {\bf 57} (1986) 957.

\bibitem{Piai:2010ma}
M.~Piai, {\it {Lectures on walking technicolor, holography and gauge/gravity
  dualities}},  {\em Adv. High Energy Phys.} {\bf 2010} (2010) 464302,
  [\href{http://arxiv.org/abs/1004.0176}{{\tt arXiv:1004.0176}}].

\bibitem{Andersen:2011yj}
J.~R. Andersen {\em et~al.}, {\it {Discovering Technicolor}},  {\em Eur. Phys.
  J. Plus} {\bf 126} (2011) 81, [\href{http://arxiv.org/abs/1104.1255}{{\tt
  arXiv:1104.1255}}].

\bibitem{Hong:2004td}
D.~K. Hong, S.~D.~H. Hsu, and F.~Sannino, {\it {Composite Higgs from higher
  representations}},  {\em Phys. Lett.} {\bf B597} (2004) 89--93,
  [\href{http://arxiv.org/abs/hep-ph/0406200}{{\tt hep-ph/0406200}}].

\bibitem{Dietrich:2006cm}
D.~D. Dietrich and F.~Sannino, {\it {Walking in the SU(N)}},  {\em Phys. Rev.}
  {\bf D75} (2007) 085018, [\href{http://arxiv.org/abs/hep-ph/0611341}{{\tt
  hep-ph/0611341}}].

\bibitem{Luty:2004ye}
M.~A. Luty and T.~Okui, {\it {Conformal technicolor}},  {\em JHEP} {\bf 09}
  (2006) 070, [\href{http://arxiv.org/abs/hep-ph/0409274}{{\tt
  hep-ph/0409274}}].

\bibitem{Azatov:2011ht}
A.~Azatov, J.~Galloway, and M.~A. Luty, {\it {Superconformal Technicolor}},
  {\em Phys. Rev. Lett.} {\bf 108} (2012) 041802,
  [\href{http://arxiv.org/abs/1106.3346}{{\tt arXiv:1106.3346}}].

\bibitem{Mack:1975je}
G.~Mack, {\it {All Unitary Ray Representations of the Conformal Group SU(2,2)
  with Positive Energy}},  {\em Commun. Math. Phys.} {\bf 55} (1977) 1.

\bibitem{Grinstein:2008qk}
B.~Grinstein, K.~A. Intriligator, and I.~Z. Rothstein, {\it {Comments on
  Unparticles}},  {\em Phys. Lett.} {\bf B662} (2008) 367--374,
  [\href{http://arxiv.org/abs/0801.1140}{{\tt arXiv:0801.1140}}].

\bibitem{Catterall:2007yx}
S.~Catterall and F.~Sannino, {\it {Minimal walking on the lattice}},  {\em
  Phys. Rev.} {\bf D76} (2007) 034504,
  [\href{http://arxiv.org/abs/0705.1664}{{\tt arXiv:0705.1664}}].

\bibitem{Catterall:2008qk}
S.~Catterall, J.~Giedt, F.~Sannino, and J.~Schneible, {\it {Phase diagram of
  SU(2) with 2 flavors of dynamical adjoint quarks}},  {\em JHEP} {\bf 11}
  (2008) 009, [\href{http://arxiv.org/abs/0807.0792}{{\tt arXiv:0807.0792}}].

\bibitem{DelDebbio:2008zf}
L.~Del~Debbio, A.~Patella, and C.~Pica, {\it {Higher representations on the
  lattice: numerical simulations. SU(2) with adjoint fermions}},  {\em Phys.
  Rev.} {\bf D81} (2010) 094503, [\href{http://arxiv.org/abs/0805.2058}{{\tt
  arXiv:0805.2058}}].

\bibitem{DelDebbio:2008tv}
L.~Del~Debbio, A.~Patella, and C.~Pica, {\it {Fermions in higher
  representations. Some results about SU(2) with adjoint fermions}},  {\em PoS}
  {\bf LATTICE2008} (2008) 064, [\href{http://arxiv.org/abs/0812.0570}{{\tt
  arXiv:0812.0570}}].

\bibitem{DelDebbio:2009fd}
L.~Del~Debbio, B.~Lucini, A.~Patella, C.~Pica, and A.~Rago, {\it {Conformal vs
  confining scenario in SU(2) with adjoint fermions}},  {\em Phys. Rev.} {\bf
  D80} (2009) 074507, [\href{http://arxiv.org/abs/0907.3896}{{\tt
  arXiv:0907.3896}}].

\bibitem{DelDebbio:2010hx}
L.~Del~Debbio, B.~Lucini, A.~Patella, C.~Pica, and A.~Rago, {\it {The infrared
  dynamics of Minimal Walking Technicolor}},  {\em Phys. Rev.} {\bf D82} (2010)
  014510, [\href{http://arxiv.org/abs/1004.3206}{{\tt arXiv:1004.3206}}].

\bibitem{DelDebbio:2010hu}
L.~Del~Debbio, B.~Lucini, A.~Patella, C.~Pica, and A.~Rago, {\it {Mesonic
  spectroscopy of Minimal Walking Technicolor}},  {\em Phys. Rev.} {\bf D82}
  (2010) 014509, [\href{http://arxiv.org/abs/1004.3197}{{\tt
  arXiv:1004.3197}}].

\bibitem{Bursa:2011ru}
F.~Bursa {\em et~al.}, {\it {Improved Lattice Spectroscopy of Minimal Walking
  Technicolor}},  {\em Phys. Rev.} {\bf D84} (2011) 034506,
  [\href{http://arxiv.org/abs/1104.4301}{{\tt arXiv:1104.4301}}].

\bibitem{Hietanen:2008mr}
A.~J. Hietanen, J.~Rantaharju, K.~Rummukainen, and K.~Tuominen, {\it {Spectrum
  of SU(2) lattice gauge theory with two adjoint Dirac flavours}},  {\em JHEP}
  {\bf 05} (2009) 025, [\href{http://arxiv.org/abs/0812.1467}{{\tt
  arXiv:0812.1467}}].

\bibitem{DelDebbio:2011kp}
L.~Del~Debbio, B.~Lucini, A.~Patella, C.~Pica, and A.~Rago, {\it {Finite volume
  effects in SU(2) with two adjoint fermions}},
  \href{http://arxiv.org/abs/1111.4672}{{\tt arXiv:1111.4672}}.

\bibitem{Appelquist:2011dp}
T.~Appelquist, G.~T. Fleming, M.~F. Lin, E.~T. Neil, and D.~A. Schaich, {\it
  {Lattice Simulations and Infrared Conformality}},  {\em Phys. Rev.} {\bf D84}
  (2011) 054501, [\href{http://arxiv.org/abs/1106.2148}{{\tt
  arXiv:1106.2148}}].

\bibitem{Hietanen:2009az}
A.~J. Hietanen, K.~Rummukainen, and K.~Tuominen, {\it {Evolution of the
  coupling constant in SU(2) lattice gauge theory with two adjoint fermions}},
  {\em Phys. Rev.} {\bf D80} (2009) 094504,
  [\href{http://arxiv.org/abs/0904.0864}{{\tt arXiv:0904.0864}}].

\bibitem{Hietanen:2009zz}
A.~Hietanen, J.~Rantaharju, K.~Rummukainen, and K.~Tuominen, {\it {Minimal
  technicolor on the lattice}},  {\em Nucl. Phys.} {\bf A820} (2009)
  191c--194c.

\bibitem{Karavirta:2011mv}
T.~Karavirta, A.~Mykkanen, J.~Rantaharju, K.~Rummukainen, and K.~Tuominen, {\it
  {Nonperturbative improvement of SU(2) lattice gauge theory with adjoint or
  fundamental flavors}},  {\em JHEP} {\bf 06} (2011) 061,
  [\href{http://arxiv.org/abs/1101.0154}{{\tt arXiv:1101.0154}}].

\bibitem{Bursa:2009tj}
F.~Bursa, L.~Del~Debbio, L.~Keegan, C.~Pica, and T.~Pickup, {\it {Running of
  the coupling and quark mass in SU(2) with two adjoint fermions}},  {\em PoS}
  {\bf LAT2009} (2009) 056, [\href{http://arxiv.org/abs/0910.2562}{{\tt
  arXiv:0910.2562}}].

\bibitem{Bursa:2009we}
F.~Bursa, L.~Del~Debbio, L.~Keegan, C.~Pica, and T.~Pickup, {\it {Mass
  anomalous dimension in SU(2) with two adjoint fermions}},  {\em Phys. Rev.}
  {\bf D81} (2010) 014505, [\href{http://arxiv.org/abs/0910.4535}{{\tt
  arXiv:0910.4535}}].

\bibitem{DeGrand:2011qd}
T.~DeGrand, Y.~Shamir, and B.~Svetitsky, {\it {Infrared fixed point in SU(2)
  gauge theory with adjoint fermions}},  {\em Phys. Rev.} {\bf D83} (2011)
  074507, [\href{http://arxiv.org/abs/1102.2843}{{\tt arXiv:1102.2843}}].

\bibitem{DeGrand:2011vp}
T.~DeGrand, Y.~Shamir, and B.~Svetitsky, {\it {Gauge theories with fermions in
  the two-index symmetric representation}},  {\em PoS} {\bf LATTICE2011} (2011)
  060, [\href{http://arxiv.org/abs/1110.6845}{{\tt arXiv:1110.6845}}].

\bibitem{Catterall:2011zf}
S.~Catterall, L.~Del~Debbio, J.~Giedt, and L.~Keegan, {\it {MCRG Minimal
  Walking Technicolor}},  {\em Phys.Rev.} {\bf D85} (2012) 094501,
  [\href{http://arxiv.org/abs/1108.3794}{{\tt arXiv:1108.3794}}].

\bibitem{Catterall:2011ce}
S.~Catterall, L.~Del~Debbio, J.~Giedt, and L.~Keegan, {\it {Systematic Errors
  of the MCRG Method}},  {\em PoS} {\bf LATTICE2011} (2011) 068,
  [\href{http://arxiv.org/abs/1110.1660}{{\tt arXiv:1110.1660}}].

\bibitem{Giedt:2012rj}
J.~Giedt and E.~Weinberg, {\it {Finite size scaling in minimal walking
  technicolor}},  {\em Phys.Rev.} {\bf D85} (2012) 097503,
  [\href{http://arxiv.org/abs/1201.6262}{{\tt arXiv:1201.6262}}].

\bibitem{Lucini:2009an}
B.~Lucini, {\it {Strongly Interacting Dynamics beyond the Standard Model on a
  Spacetime Lattice}},  {\em Phil. Trans. Roy. Soc. Lond.} {\bf A368} (2010)
  3657--3670, [\href{http://arxiv.org/abs/0911.0020}{{\tt arXiv:0911.0020}}].

\bibitem{Patella:2011jr}
A.~Patella, {\it {GMOR-like relation in IR-conformal gauge theories}},  {\em
  Phys. Rev.} {\bf D84} (2011) 125033,
  [\href{http://arxiv.org/abs/1106.3494}{{\tt arXiv:1106.3494}}].

\bibitem{DeGrand:2009et}
T.~DeGrand, {\it {Volume scaling of Dirac eigenvalues in SU(3) lattice gauge
  theory with color sextet fermions}},
  \href{http://arxiv.org/abs/0906.4543}{{\tt arXiv:0906.4543}}.

\bibitem{DelDebbio:2010ze}
L.~Del~Debbio and R.~Zwicky, {\it {Hyperscaling relations in mass-deformed
  conformal gauge theories}},  {\em Phys. Rev.} {\bf D82} (2010) 014502,
  [\href{http://arxiv.org/abs/1005.2371}{{\tt arXiv:1005.2371}}].

\bibitem{Cheng:2011ic}
A.~Cheng, A.~Hasenfratz, and D.~Schaich, {\it {Novel phase in SU(3) lattice
  gauge theory with 12 light fermions}},  {\em Phys.Rev.} {\bf D85} (2012)
  094509, [\href{http://arxiv.org/abs/1111.2317}{{\tt arXiv:1111.2317}}].

\bibitem{Giusti:2008vb}
L.~Giusti and M.~Luscher, {\it {Chiral symmetry breaking and the Banks--Casher
  relation in lattice QCD with Wilson quarks}},  {\em JHEP} {\bf 03} (2009)
  013, [\href{http://arxiv.org/abs/0812.3638}{{\tt arXiv:0812.3638}}].

\bibitem{DelDebbio:2005qa}
L.~Del~Debbio, L.~Giusti, M.~Luscher, R.~Petronzio, and N.~Tantalo, {\it
  {Stability of lattice QCD simulations and the thermodynamic limit}},  {\em
  JHEP} {\bf 0602} (2006) 011,
  [\href{http://arxiv.org/abs/hep-lat/0512021}{{\tt hep-lat/0512021}}].

\end{thebibliography}\endgroup

\begin{table}[!p]
\centering
\parbox[b]{6cm}{
\centering
\begin{tabular}{|c|c|}
\hline
$a \Omega$ & $ a^{-4} \bar{\nu}(\Omega) $ \\
\hline
\hline
$ 0.0800 $ & $ 1.31(42) \times 10^{-6}  $ \\ \hline
$ 0.0812 $ & $ 2.62(64) \times 10^{-6}  $ \\ \hline
$ 0.0824 $ & $ 0.64(11) \times 10^{-5}  $ \\ \hline
$ 0.0836 $ & $ 1.07(12) \times 10^{-5}  $ \\ \hline
$ 0.0848 $ & $ 1.52(13) \times 10^{-5}  $ \\ \hline
$ 0.0860 $ & $ 2.05(12) \times 10^{-5}  $ \\ \hline
$ 0.0872 $ & $ 2.73(14) \times 10^{-5}  $ \\ \hline
$ 0.0884 $ & $ 3.39(16) \times 10^{-5}  $ \\ \hline
$ 0.0896 $ & $ 4.10(20) \times 10^{-5}  $ \\ \hline
$ 0.0908 $ & $ 4.99(19) \times 10^{-5}  $ \\ \hline
$ 0.0920 $ & $ 5.91(21) \times 10^{-5}  $ \\ \hline
$ 0.0932 $ & $ 6.84(23) \times 10^{-5}  $ \\ \hline
$ 0.0944 $ & $ 7.78(24) \times 10^{-5}  $ \\ \hline
$ 0.0956 $ & $ 8.94(26) \times 10^{-5}  $ \\ \hline
$ 0.0968 $ & $ 1.007(27) \times 10^{-4} $ \\ \hline
$ 0.0980 $ & $ 1.122(28) \times 10^{-4} $ \\ \hline
$ 0.0992 $ & $ 1.252(28) \times 10^{-4} $ \\ \hline
$ 0.1004 $ & $ 1.390(30) \times 10^{-4} $ \\ \hline
$ 0.1016 $ & $ 1.520(31) \times 10^{-4} $ \\ \hline
$ 0.1028 $ & $ 1.669(31) \times 10^{-4} $ \\ \hline
$ 0.1040 $ & $ 1.812(31) \times 10^{-4} $ \\ \hline
$ 0.1052 $ & $ 1.972(30) \times 10^{-4} $ \\ \hline
$ 0.1064 $ & $ 2.141(33) \times 10^{-4} $ \\ \hline
$ 0.1111 $ & $ 2.858(34) \times 10^{-4} $ \\ \hline
\end{tabular}
}\parbox[b]{6cm}{
\centering
\begin{tabular}{|c|c|}
\hline
$a \Omega$ & $ a^{-4} \bar{\nu}(\Omega) $ \\
\hline
\hline
$ 0.1163 $ & $ 3.648(40) \times 10^{-4} $ \\ \hline
$ 0.1217 $ & $ 4.767(55) \times 10^{-4} $ \\ \hline
$ 0.1274 $ & $ 6.027(50) \times 10^{-4} $ \\ \hline
$ 0.1333 $ & $ 7.495(54) \times 10^{-4} $ \\ \hline
$ 0.1395 $ & $ 9.144(73) \times 10^{-4} $ \\ \hline
$ 0.1460 $ & $ 1.1093(73) \times 10^{-3} $ \\ \hline
$ 0.1529 $ & $ 1.3622(67) \times 10^{-3} $ \\ \hline
$ 0.1600 $ & $ 1.6221(79) \times 10^{-3} $ \\ \hline
$ 0.1674 $ & $ 1.9448(67) \times 10^{-3} $ \\ \hline
$ 0.1753 $ & $ 2.3011(99) \times 10^{-3} $ \\ \hline
$ 0.1834 $ & $ 2.7354(95) \times 10^{-3} $ \\ \hline
$ 0.2009 $ & $ 3.816(12) \times 10^{-3} $ \\ \hline
$ 0.2201 $ & $ 5.282(13) \times 10^{-3} $ \\ \hline
$ 0.2411 $ & $ 7.217(16) \times 10^{-3} $ \\ \hline
$ 0.2641 $ & $ 9.810(19) \times 10^{-3} $ \\ \hline
$ 0.2894 $ & $ 1.3206(18) \times 10^{-2} $ \\ \hline
$ 0.3170 $ & $ 1.7659(23) \times 10^{-2} $ \\ \hline
$ 0.3472 $ & $ 2.3649(31) \times 10^{-2} $ \\ \hline
$ 0.3804 $ & $ 3.1452(29) \times 10^{-2} $ \\ \hline
$ 0.4167 $ & $ 4.1708(47) \times 10^{-2} $ \\ \hline
$ 0.4564 $ & $ 5.5120(43) \times 10^{-2} $ \\ \hline
$ 0.5000 $ & $ 7.2580(61) \times 10^{-2} $ \\ \hline
$ 0.5477 $ & $ 9.5130(62) \times 10^{-2} $ \\ \hline
$ 0.6000 $ & $ 1.24498(63) \times 10^{-1} $ \\ \hline
\end{tabular}
}
\caption{Set S1 ($64\times 24^3$ $\beta=2.25$ $am_0=-1.15$). Mode number per unit volume, computed from the eigenvalues for $a\Omega < 0.11$ and with the projector method for $a\Omega > 0.11$.}
\label{tab:S1}
\end{table}

\begin{table}
\centering
\begin{tabular}{|c||c|c|}
\hline
$a \Omega$ & $a^{-4} \bar{\nu}(\Omega)$ @ $64 \times 24^3$ (S1) & $a^{-4} \bar{\nu}(\Omega)$ @ $64 \times 32^3$ (S2) \\
\hline
\hline
$0.086$ & $2.05(12) \times 10^{-5}$ & $1.974(65) \times 10^{-5}$ \\ \hline
$0.092$ & $5.91(21) \times 10^{-5}$ & $5.90(23) \times 10^{-5}$ \\ \hline
$0.098$ & $1.1122(28) \times 10^{-4}$ & $1.1105(26) \times 10^{-4}$ \\ \hline
$0.104$ & $1.811(31) \times 10^{-4}$ & $1.827(19) \times 10^{-4}$ \\ \hline
\end{tabular}
\caption{The mode number for the set S2 ($64\times 32^3$ $\beta=2.25$ $am_0=-1.15$), computed with the projector method, is compared to the mode number for selected points of the set S1 ($64\times 24^3$ $\beta=2.25$ $am_0=-1.15$). For all the considered eigenvalues, the finite-volume effects are negligible.}
\label{tab:finitevolume}
\end{table}

\begin{table}
\centering
\begin{tabular}{|c|c|}
\hline
$a \Omega$ & $ a^{-4} \bar{\nu}(\Omega) $ \\
\hline
\hline
$0.092$ & $3.567(52) \times 10^{-4}$ \\ \hline
$0.098$ & $4.510(62) \times 10^{-4}$ \\ \hline
$0.104$ & $5.473(51) \times 10^{-4}$ \\ \hline
$0.110$ & $6.834(60) \times 10^{-4}$ \\ \hline
$0.116$ & $7.997(72) \times 10^{-4}$ \\ \hline
$0.122$ & $9.508(66) \times 10^{-4}$ \\ \hline
$0.128$ & $1.123(87) \times 10^{-3}$ \\ \hline
$0.134$ & $1.319(12) \times 10^{-3}$ \\ \hline
$0.140$ & $1.5146(77) \times 10^{-3}$ \\ \hline
$0.150$ & $1.896(11) \times 10^{-3}$ \\ \hline
$0.160$ & $2.337(12) \times 10^{-3}$ \\ \hline
$0.180$ & $3.423(16) \times 10^{-3}$ \\ \hline
\end{tabular}
\caption{Set S3 ($64\times 24^3$ $\beta=2.25$ $am_0=-1.18$). Mode number per unit volume, computed with the projector method.}
\label{tab:S3}
\end{table}

\begin{table}
\hspace{-10mm} \small
\begin{tabular}{|c||c||c|c|c|c||c|c|}
\hline
Fit & Range & $a^{-4} \bar{\nu}_0$ & $A$ & $am$ & $\gamma_*$ & dof & $\chi^2/\textrm{dof}$ \\
\hline
\hline
S1:F1 & $0.08 \le a\Omega \le 0.18$	&	$2.00(75) \times 10^{-6}$	&	$0.586(12)$	&	$8.03(32) \times 10^{-2}$	&	$0.3431(52)$	&	$31$	&	$0.68$ \\ \hline
S1:F2 & $0.083 \le a\Omega \le 0.18$	&	$0.53(18) \times 10^{-5}$	&	$0.560(18)$	&	$8.11(52) \times 10^{-2}$	&	$0.3556(86)$	&	$28$	&	$0.24$ \\ \hline
S1:F3 & $0.087 \le a\Omega \le 0.18$	&	$0.67(35) \times 10^{-5}$	&	$0.553(23)$	&	$8.14(85) \times 10^{-2}$	&	$0.359(12)$	&	$25$	&	$0.27$ \\ \hline
\textbf{S1:F4} & $\bm{0.091 \le a\Omega \le 0.18}$	&	$\bm{1.31(78) \times 10^{-5}}$	&	$\bm{0.532(38)}$	&	$\bm{8.26(16) \times 10^{-2}}$	&	$\bm{0.371(20)}$	&	$\bm{21}$	&	$\bm{0.30}$ \\ \hline 
S1:F5 & $0.094 \le a\Omega \le 0.18$	&	$1.46(98) \times 10^{-5}$	&	$0.528(41)$	&	$8.28(19) \times 10^{-2}$	&	$0.373(23)$	&	$19$	&	$0.33$ \\ \hline
S1:F6 & $0.096 \le a\Omega \le 0.18$	&	$0.17(12) \times 10^{-4}$	&	$0.524(43)$	&	$8.32(21) \times 10^{-2}$	&	$0.376(24)$	&	$17$	&	$0.36$ \\ \hline
S1:F7 & $0.1 \le a\Omega \le 0.18$	&	$0.19(20) \times 10^{-4}$	&	$0.519(53)$	&	$8.35(33) \times 10^{-2}$	&	$0.378(31)$	&	$14$	&	$0.43$ \\ \hline
\hline
S1:F8 & $0.091 \le a\Omega \le 0.105$	&	$0.05(24) \times 10^{-4}$	&	$0.58(31)$	&	$8.09(56) \times 10^{-2}$	&	$0.35(15)$	&	$7$	&	$0.024$\\ \hline 
S1:F9 & $0.091 \le a\Omega \le 0.115$	&	$-0.08(23) \times 10^{-4}$	&	$1.0(1.8)$	&	$7.65(90) \times 10^{-2}$	&	$0.24(15)$	&	$11$	&	$0.17$\\ \hline 
S1:F10 & $0.091 \le a\Omega \le 0.12$	&	$0.15(23) \times 10^{-4}$	&	$0.47(60)$	&	$8.33(65) \times 10^{-2}$	&	$0.40(14)$	&	$12$	&	$0.21$\\ \hline 
S1:F11 & $0.091 \le a\Omega \le 0.125$	&	$-0.02(23) \times 10^{-4}$	&	$0.72(70)$	&	$7.88(77) \times 10^{-2}$	&	$0.30(12)$	&	$13$	&	$0.21$\\ \hline 
S1:F12 & $0.091 \le a\Omega \le 0.13$	&	$-0.03(31) \times 10^{-4}$	&	$0.73(83)$	&	$0.79(11) \times 10^{-1}$	&	$0.30(12)$	&	$14$	&	$0.19$\\ \hline 
S1:F13 & $0.091 \le a\Omega \le 0.135$	&	$0.10(21) \times 10^{-4}$	&	$0.57(30)$	&	$8.18(57) \times 10^{-2}$	&	$0.354(78)$	&	$15$	&	$0.20$\\ \hline 
S1:F14 & $0.091 \le a\Omega \le 0.14$	&	$0.19(11) \times 10^{-4}$	&	$0.490(83)$	&	$8.38(25) \times 10^{-2}$	&	$0.392(40)$	&	$16$	&	$0.22$\\ \hline 
S1:F15 & $0.091 \le a\Omega \le 0.15$	&	$2.44(100) \times 10^{-5}$	&	$0.449(68)$	&	$8.50(21) \times 10^{-2}$	&	$0.415(37)$	&	$17$	&	$0.22$\\ \hline 
S1:F16 & $0.091 \le a\Omega \le 0.16$	&	$0.13(10) \times 10^{-4}$	&	$0.533(59)$	&	$8.25(22) \times 10^{-2}$	&	$0.370(29)$	&	$19$	&	$0.33$\\ \hline 
S1:F17 & $0.091 \le a\Omega \le 0.17$	&	$1.26(83) \times 10^{-5}$	&	$0.535(41)$	&	$8.25(16) \times 10^{-2}$	&	$0.369(22)$	&	$20$	&	$0.31$\\ \hline 
S1:F18 & $0.091 \le a\Omega \le 0.19$	&	$0.76(81) \times 10^{-5}$	&	$0.561(36)$	&	$8.14(16) \times 10^{-2}$	&	$0.356(19)$	&	$22$	&	$0.33$\\ \hline 
S1:F19 & $0.091 \le a\Omega \le 0.21$	&	$0.38(72) \times 10^{-5}$	&	$0.579(29)$	&	$8.07(14) \times 10^{-2}$	&	$0.347(15)$	&	$23$	&	$0.33$\\ \hline 
S1:F20 & $0.091 \le a\Omega \le 0.23$	&	$-0.57(61) \times 10^{-5}$	&	$0.620(21)$	&	$7.87(12) \times 10^{-2}$	&	$0.327(11)$	&	$24$	&	$0.45$\\ \hline 
S1:F21 & $0.091 \le a\Omega \le 0.25$	&	$-0.82(54) \times 10^{-5}$	&	$0.630(15)$	&	$7.82(10) \times 10^{-2}$	&	$0.3223(78)$	&	$25$	&	$0.45$\\ \hline 
S1:F22 & $0.091 \le a\Omega \le 0.27$	&	$-1.19(51) \times 10^{-5}$	&	$0.641(13)$	&	$7.74(94) \times 10^{-2}$	&	$0.3166(66)$	&	$26$	&	$0.50$\\ \hline 
S1:F23 & $0.091 \le a\Omega \le 0.29$	&	$-0.85(41) \times 10^{-5}$	&	$0.6326(75)$	&	$7.81(72) \times 10^{-2}$	&	$0.3212(42)$	&	$27$	&	$0.52$\\ \hline 
S1:F24 & $0.091 \le a\Omega \le 0.39$	&	$0.26(38) \times 10^{-5}$	&	$0.6122(27)$	&	$8.01(68) \times 10^{-2}$	&	$0.3331(21)$	&	$30$	&	$0.84$\\ \hline 
S1:F25 & $0.091 \le a\Omega \le 0.5$	&	$1.64(35) \times 10^{-5}$	&	$0.5968(15)$	&	$8.23(61) \times 10^{-2}$	&	$0.3444(15)$	&	$33$	&	$2.4$\\ \hline 
S1:F26 & $0.091 \le a\Omega \le 0.6$	&	$3.58(52) \times 10^{-5}$	&	$0.5805(13)$	&	$8.54(83) \times 10^{-2}$	&	$0.3584(18)$	&	$35$	&	$10$\\ \hline 
\hline
S3:F1 & $0.091 \le a\Omega \le 0.13$	&	$-0.04(16) \times 10^{-3}$	&	$0.85(31)$	&	$0.12(45) \times 10^{-1}$	&	$0.25(18)$ &	$6$	&	$1.4$ \\ \hline 
S3:F2 & $0.091 \le a\Omega \le 0.14$	&	$0.02(11) \times 10^{-3}$	&	$0.81(30)$	&	$0.37(34) \times 10^{-1}$	&	$0.27(14)$	&	$8$	&	$1.1$ \\ \hline 
S3:F3 & $0.091 \le a\Omega \le 0.15$	&	$0.06(12) \times 10^{-3}$	&	$0.67(24)$	&	$0.50(33) \times 10^{-1}$	&	$0.33(12)$	&	$9$	&	$1.0$ \\ \hline 
S3:F4 & $0.091 \le a\Omega \le 0.16$	&	$0.10(10) \times 10^{-3}$	&	$0.59(17)$	&	$0.57(27) \times 10^{-1}$	&	$0.364(92)$	&	$10$	&	$0.95$ \\ \hline 
\textbf{S3:F5} & $\bm{0.091 \le a\Omega \le 0.18}$	&	$\bm{0.54(73) \times 10^{-4}}$	&	$\bm{0.667(98)}$	&	$\bm{0.48(22) \times 10^{-1}}$	&	$\bm{0.325(50)}$	&	$\bm{11}$	&	$\bm{0.87}$ \\ \hline 
\end{tabular}
\caption{Eq.~\eqref{eq:nu_powerlaw_lattice} is fitted to the mode number per unit volume for sets S1 and S3. The fit range for set S1 ($am_0=-1.15$) is systematically explored. In fits S1:F1 to S1:F7, the fit-range higher end is kept fixed at $0.18$ while the lower end is varied. In fits S1:F8 to S1:F26, the lower end is kept fixed at $0.091$ while the higher end varied. S1:F4 (in bold) has been chosen as the final result. Fits S3:F1 to S3:F5 correspond to the lighter mass $am_0=-1.18$. $\gamma_*$ obtained from this set is compatible with the heavier-mass result.}
\label{tab:fits}
\end{table}

\end{document}